\documentclass[12pt]{iopart}

\usepackage{iopams}
\usepackage{setstack}
\usepackage{graphicx}
\usepackage{epsfig}
\usepackage{url}
\usepackage{epstopdf}
\usepackage{hyperref}

\newcommand{\be}{\begin{equation}}
\newcommand{\ee}{\end{equation}}

\newcommand{\bea}{\begin{eqnarray}}
\newcommand{\eea}{\end{eqnarray}}
\newcommand{\bean}{\begin{eqnarray*}}
\newcommand{\eean}{\end{eqnarray*}}

\begin{document}

\title{A Rule of Thumb for the Detectability of Gravitational-Wave Bursts}
\author{Patrick J.~Sutton}
\address{School of Physics and Astronomy, Cardiff University, Cardiff, United Kingdom, \mbox{CF24 3AA}}
\ead{patrick.sutton@astro.cf.ac.uk} 

\begin{abstract}
We derive a simple relationship between the energy emitted in gravitational waves 
for a narrowband source and the distance to which that emission can be detected 
by a single detector.  
We consider linearly polarized, elliptically polarized, and unpolarized  gravitational waves, and emission patterns appropriate for each of these 
cases.  We ignore cosmological effects.
\end{abstract}


\section{Introduction}

The sensitivity of a gravitational-wave detector to a transient signal (``burst'') 
is reasonably well characterised by the expectation value of the matched-filter 
signal-to-noise ratio (SNR) of the burst \cite{AnBrCrFl:01}.  
This expectation value can be computed as an integral of the burst 
signal spectrum divided by the detector noise spectral density.  
For the most commonly considered gravitational-wave transient signals, 
the inspiral of neutron star binaries, this leads to the well-known 
average (or ``sensemon''\footnote{
The other commonly used range for inspirals is the ``horizon'' range, 
defined as the maximum distance at which an optimally positioned and 
oriented binary would produce an expected SNR of at least $\rho_\mathrm{det}$.  The horizon range is a factor 2.26 larger than the average range.
}) range \cite{FiCh:93} giving the effective volume within which the expected SNR in a single detector would be 
above some threshold $\rho_\mathrm{det}$:
\begin{equation}\label{eqn:sensemon}
{\cal R}_\mathrm{BNS} = 
1.77
\left(\frac{5{\cal M}^{5/3}f_{7/3}}{96\pi^{4/3}\rho_\mathrm{det}^2}\right)^{1/2} \, .
\end{equation}
Here ${\cal M}$ is the chirp mass of the binary and $f_{7/3}=\int df f^{-7/3} S^{-1}$ where $S$ is the noise power spectrum of the detector.  The factor $1.77$ arises from integration over possible positions and orientations of the binary.
The average range ${\cal R}_\mathrm{BNS}$ is defined such that 
for a homogenous isotropic distribution of sources with rate density 
$\dot{{\cal N}}$ the mean rate of detections will then be 
\begin{equation}
\dot{N} = \frac{4}{3}\pi ({\cal R}_\mathrm{BNS})^3 \dot{{\cal N}} \, .
\end{equation}

The average range is clearly a useful tool for assessing the scientific 
capabilities of a detector.
Unfortunately, while the expected gravitational-wave emission of binary 
systems is well-understood, gravitational wave transients from most other 
potential sources such as supernovae, long gamma-ray bursts, and soft gamma 
repeaters are not known reliably.  
In this technical note we present a simple definition of the effective 
sensitive range of a gravitational-wave detector to generic bursts, based 
on the total energy emitted in the burst and the peak frequency of the emission. 
This range is appropriate for a variety of possible signal polarizations and 
source emission patterns, and requires only that the signal bandwidth is smaller
than the frequency range over which the detector noise spectrum varies significantly.  

We begin by determining how the total energy $E_\mathrm{GW}$ carried in 
a gravitational-wave burst is related to the measure of signal strength 
commonly used in burst searches, the root-sum-square amplitude 
$h_\mathrm{rss}$.  We then relate these measures to the expected 
signal-to-noise ratio $\rho$, and 
define an average range analogous to equation (\ref{eqn:sensemon}).  
Finally, we compare this range to the results of recent LIGO-Virgo 
searches for gravitational-wave bursts \cite{S5y1,S5y2,Ab_etal:12}.

\section{Relating $E_\mathrm{GW}$ to $h_\mathrm{rss}$}
\label{sec:intro}

We first relate the total energy emitted in gravitational waves, $E_\mathrm{GW}$, 
to the LIGO-Virgo standard measure for burst amplitude at the detector, $h_\mathrm{rss}$.  The latter is defined by
\begin{eqnarray}
h_\mathrm{rss} 
 & = & \int_{-\infty}^\infty \!dt \, \left[ h^2_+(t) + h_\times^2(t) \right] \\
 & = & 2 \int_0^\infty \!df \, \left[ |\tilde{h}_+(f)|^2 + |\tilde{h}_\times(f)|^2 \right]
   \, .
\end{eqnarray}
The flux (energy per unit area per unit time) of a gravitational wave is 
\begin{equation}
F_\mathrm{GW} = \frac{c^3}{16 \pi G} \langle \dot{h}_+^2(t) + \dot{h}_\times^2(t) \rangle \, ,
\end{equation}
where the angle brackets denote an average over several periods.
For a burst of duration $\le T$ we can compute the average by integrating over the duration:
\begin{eqnarray}
\fl 
~~~~~~~~~~~F_\mathrm{GW} 
  & = &  \frac{c^3}{16 \pi G} \frac{1}{T} \int_{-T/2}^{T/2} \!\!\!dt\, \left[ \dot{h}_+^2(t) + \dot{h}_\times^2(t) \right] \\
  & = &  \frac{c^3}{16 \pi G} \frac{1}{T} \int_{-T/2}^{T/2} \!\!\!dt\, 
             \left[ \vphantom{\int_{-\infty}^\infty} \right.  
             \int_{-\infty}^\infty \!\!\!df' \e^{i2\pi f't}(i2\pi f') \tilde{h}_+^*(f') \,
             \int_{-\infty}^\infty \!\!\!df  \e^{-i2\pi ft}(-i2\pi f) \tilde{h}_+(f) 
             \nonumber \\
  &   &    \left. \vphantom{\int_{-\infty}^\infty}    
           \qquad + (\mathrm{same,}~+\to\times)\right] 
\end{eqnarray}
Since $h_{+,\times}\to0$ outside $-T/2 < t < T/2$, we may extend the time 
integration to $t\to\pm\infty$. The time integral then evaluates to a delta
function, $\delta(f-f')$, giving 
\begin{equation}
F_\mathrm{GW} 
  =  \frac{\pi c^3}{4 G} \frac{1}{T} \int_{-\infty}^\infty \!\!\!df f^2 \left( 
         |\tilde{h}_+(f)|^2 + |\tilde{h}_\times(f)|^2 \right) \, .
\end{equation}

\subsection{Isotropic emission}

To compute the total energy $E_\mathrm{GW}$ emitted, we need to integrate the 
flux $F_\mathrm{GW}$ assuming some emission pattern.  Let us first assume 
isotropic emission.  Then for a source at a distance $r$
\begin{eqnarray}
E_\mathrm{GW} 
  & = &  4\pi r^2 \, T F_\mathrm{GW} \\
  & = &  \frac{\pi^2 c^3}{G} r^2 \int_{-\infty}^\infty \!\!\!df f^2 \left( 
         |\tilde{h}_+(f)|^2 + |\tilde{h}_\times(f)|^2 \right) \, .
\end{eqnarray}
If we assume that the signal is narrowband with central frequency $f_0$, we obtain 
\begin{equation}
\label{eqn:Eiso}
E_\mathrm{GW} 
  =  \frac{\pi^2 c^3}{G} r^2 f_0^2 h_\mathrm{rss}^2 \, .
\end{equation}

\subsection{Linear motion emission}

Axisymmetric motion will produce linearly polarized emission with pattern 
\begin{eqnarray}
  h_+(t)      & = & \sin^2(\iota) \, h(t) \, , \\
  h_\times(t) & = & 0 \, ,
\end{eqnarray}
where $\iota$ is the angle between the symmetry axis and the line-of-sight 
to the observer, and we have selected a polarization basis aligned with 
this symmetry axis.  The energy emitted in a narrowband signal is then 
\begin{eqnarray}
E_\mathrm{GW} 
  & = &  \frac{\pi c^3}{4G} r^2 \,
         \int_{-1}^{1}\!\!\!d(\cos\iota) \int_0^{2\pi}\!\!\!d\lambda \,
         \int_{-\infty}^\infty \!\!\!df f^2 \left( 
         \sin^4(\iota) \, |\tilde{h}(f)|^2 \right) \nonumber \\
  & = &  \frac{8}{15} \frac{\pi^2 c^3}{G} r^2 f_0^2 h_\mathrm{rss}^2 
         \, , \label{eqn:Elinear}
\end{eqnarray}
where $\lambda$ is the azimuthal angle in the source frame.
This is $8/15$ times the result for isotropic emission, (\ref{eqn:Eiso}).

Note that in writing (\ref{eqn:Elinear}) we have defined 
$h_\mathrm{rss}$ as the root-sum-square amplitude 
from an {\em optimally oriented} source ($\iota=\pi/2$ in this case).  This differs slightly from the 
standard LIGO-Virgo definition, which includes the inclination factors.  
In practice, however, LIGO-Virgo papers 
to date have typically simulated optimally oriented sources.

\subsection{Rotating system emission}

Rotational motion (such as from a circular binary) will produce emission with pattern 
\begin{eqnarray}
  h_+(t)      & = & \frac12(1+\cos^2(\iota)) \, A(t) \cos\Phi(t) \, , \\
  h_\times(t) & = & \cos(\iota) \, A(t) \sin\Phi(t) \, ,
\end{eqnarray}
where $\iota$ is the angle between the rotation axis and the line-of-sight 
to the observer, and we have again selected a polarization basis aligned with 
this symmetry axis.  We assume $A(t)$ varies slowly enough compared to 
$\Phi(t)$ that $h_+$ and $h_\times$ are approximately orthogonal.  
This produces an elliptically polarized signal at the detector.
The energy emitted in a narrowband signal is  
\begin{eqnarray}
E_\mathrm{GW} 
  & = &  \frac{\pi c^3}{4G} r^2 \,
         \int_{-1}^{1}\!\!\!d(\cos\iota) \int_0^{2\pi}\!\!\!d\lambda \,
         \int_{-\infty}^\infty \!\!\!df f^2 \left( 
             \frac{(1+\cos^2(\iota))^2}{4} + \cos^2(\iota) 
         \right) \, |\tilde{h}(f)|^2 \nonumber \\
  & = &  \frac{2}{5} \frac{\pi^2 c^3}{G} r^2 f_0^2 h_\mathrm{rss}^2 
         \, , \label{eqn:Eell}
\end{eqnarray}
where $\tilde{h}(f)$ is the Fourier transform of $A(t) \cos\Phi(t)$ 
and we have again used $h_\mathrm{rss}$ for an optimally oriented 
source ($\iota=0$).
The expression for energy emitted is $2/5$ times the result for 
isotropic emission, (\ref{eqn:Eiso}).

\section{Relating $E_\mathrm{GW}$ to Signal-To-Noise Ratio}

The detectability of a generic signal is determined mainly by its 
expected signal-to-noise ratio $\rho$ for a matched filter.   
(The time-frequency volume $V_\mathrm{TF}$ of the signal is also important 
when $V_\mathrm{TF}\gg1$ \cite{AnBrCrFl:01,Su_etal:10}).  For a 
narrowband signal, $\rho$ has a simple relationship to the 
$h_\mathrm{rss}$ amplitude.  We start from 
\begin{eqnarray}\label{eqn:SNR}
\rho^2 
  & = &  2 \int_{-\infty}^\infty \!\!\!\!\!df \, \frac{|F_+\tilde{h}_+(f) + F_\times\tilde{h}_\times(f)|^2}{S(f)} \, ,
\end{eqnarray}
where $S(f)$ is the one-sided noise power spectrum, and 
$F_{+,\times}(\theta,\phi,\psi)$ are the antenna responses to the sky 
position $(\theta,\phi)$ and polarization $\psi$ of the gravitational wave.  
We may expand the square in (\ref{eqn:SNR}) and drop the 
$\tilde{h}_+ \tilde{h}_\times^*$ terms for most signals of interest: 
for elliptically polarized signals the two waveforms are orthogonal, while for 
linearly polarized signals $\tilde{h}_\times=0$.  The waveforms are also 
orthogonal in the {\em unpolarized} case, where the two polarizations are 
independent stochastic timeseries. An example is white-noise bursts \cite{Ab_etal:12}.
Assuming a narrowband signal, we find
\begin{eqnarray}\label{eqn:SNR2}
\rho^2 
  & = &  \Theta^2 \frac{h_\mathrm{rss}^2}{S(f_0)} \, ,
\end{eqnarray}
where we define the angle factor
\begin{eqnarray}\label{eqn:Theta}
\fl
~~~~\Theta^2 \equiv  \left\{ \begin{array}{l l} 
            F_+^2(\theta,\phi,\psi) + F_\times^2(\theta,\phi,\psi) & ~\mathrm{isotropic~unpolarized} \\
            F_+^2(\theta,\phi,\psi) (\frac{1+\cos^2(\iota)}{2})^2 + F_\times^2(\theta,\phi,\psi) \cos^2(\iota) & ~\mathrm{elliptical} \\
            F_+^2(\theta,\phi,\psi) \, 2 \sin^4\iota \vphantom{{\sin^4}^1}  & ~\mathrm{linear}
        \end{array} \right.
\end{eqnarray}
Note that all dependence on the four angles $\theta$, $\phi$, $\psi$, and $\iota$ 
is contained in $\Theta$.  Substituting (\ref{eqn:Eiso}), 
(\ref{eqn:Elinear}), or (\ref{eqn:Eell}) gives  
\begin{equation}\label{eqn:SNR3}
\rho^2 
  =  \frac{\Theta^2}{\alpha} \frac{G}{\pi^2 c^3} \frac{E_\mathrm{GW}}{S(f_0) r^2 f_0^2} \, ,
\end{equation}
where $\alpha=1$ for isotropic emission, $8/15$ for linearly polarized 
emission, and $2/5$ for circularly polarized emission.

\section{Effective Range}

We can now combine the results for $E_\mathrm{GW}$ and $\rho$ to compute 
the typical distance to which a source is detectable.  We will follow the 
approach used in Section V of \cite{FiCh:93}.

Consider a homogenous isotropic distribution of sources with rate density 
$\dot{{\cal N}}$.  A signal from a given source will be detectable if the 
received signal-to-noise is above some threshold value $\rho_\mathrm{det}$.  
The mean rate of detections will then be 
\begin{equation}
\dot{N} 
  =  4\pi \dot{{\cal N}} \int_0^\infty dr r^2 P(\rho^2 > \rho_\mathrm{det}^2) \, .
\end{equation}
Here $P(\rho^2 > \rho_\mathrm{det}^2)$ is the probability that the signal-to-noise 
of a source at given distance $r$ with random $\theta$, $\phi$, $\psi$, and 
$\iota$ will be above threshold.  Using (\ref{eqn:SNR3}), we may write this 
probability as
\begin{eqnarray}
P(\rho^2 > \rho_\mathrm{det}^2)
  & = &  P(\Theta^2 > r^2/r_0^2) \, ,
\end{eqnarray}
where we have defined the fiducial distance 
\begin{equation}
  r_0^2 = \frac{G}{\alpha \pi^2 c^3} \frac{E_\mathrm{GW}}{S(f_0) f_0^2 \rho_\mathrm{det}^2} \, .
\end{equation}
Our detection rate is thus
\begin{equation}
\dot{N}_\mathrm{det} 
  =  \frac{4}{3}\pi r_0^3 \dot{N} \left[ 3 \int_0^\infty\!\!\!dx\,x^2 P(\Theta^2 > x^2) \right] \, .
\end{equation}
The integral is easily evaluated numerically:
\begin{eqnarray}\label{eqn:integral}
\int_0^\infty\!\!\!dx\,x^2 P(\Theta^2 > x^2) 
  =  \left\{ \begin{array}{l l} 
            0.0978 & ~\mathrm{unpolarized} \\
            0.0287 & ~\mathrm{elliptical} \\
            0.0537 & ~\mathrm{linear}
        \end{array} \right.
\end{eqnarray}
Following \cite{FiCh:93}, we define the effective detection range 
${\cal R}_\mathrm{eff}$ as the radius enclosing a spherical 
volume $V$ such that the rate of detections is $\dot{{\cal N}} V$:
\begin{eqnarray}
{\cal R}_\mathrm{eff} 
  & = &  r_0 \left[ 3 \int_0^\infty\!\!\!dx\,x^2 P(\Theta^2 > x^2) \right]^{1/3} \\
  & = &  \beta \left(\frac{G}{\pi^2 c^3} \frac{E_\mathrm{GW}}{S(f_0) f_0^2 \rho_\mathrm{det}^2}\right)^{1/2} \, ,
\end{eqnarray}
where 
\begin{eqnarray}\label{eqn:beta}
\beta \equiv \alpha^{-1/2} \left[ 3 \int_0^\infty\!\!\!dx\,x^2 P(\Theta^2 > x^2) \right]^{1/3} 
  & = &  \left\{ \begin{array}{l l} 
             0.665 & ~\mathrm{unpolarized} \\
             0.698 & ~\mathrm{elliptical} \\
             0.745 & ~\mathrm{linear}
         \end{array} \right. \, .
\end{eqnarray}
We note that for all three cases (unpolarized, linear, and elliptical polarizations), $\beta$ is equal to $1/\sqrt{2}$ to within a few percent.  A convenient approximation is thus 
\begin{eqnarray}\label{eqn:range}
{\cal R}_\mathrm{eff} 
  & \simeq &  \left(\frac{G}{2\pi^2 c^3} \frac{E_\mathrm{GW}}{S(f_0) f_0^2 \rho_\mathrm{det}^2}\right)^{1/2} \, .
\end{eqnarray}
With this definition the mean rate of detections 
for a homogenous isotropic distribution of standard-candle (fixed $E_\mathrm{GW}, f_0$) burst sources with rate density 
$\dot{{\cal N}}$ is 
\begin{equation}\label{eqn:rate}
\dot{N} = \frac{4}{3}\pi ({\cal R}_\mathrm{eff})^3 \dot{{\cal N}} \, .
\end{equation}

\section{Example: LIGO-Virgo Science Runs, 2005--2010}

As an example, we apply our effective range formula (\ref{eqn:range}) to 
the LIGO-Virgo network during their 2005--07 and 2009--10 science runs.  
The results of the search for generic gravitational-wave bursts are reported 
in \cite{S5y1,S5y2,Ab_etal:12}.
Approximately 1.8\,yr of coincident data were analysed from the three LIGO 
detectors (H1, H2, L1) and the Virgo detector (V1).  No gravitational waves were 
detected, and limits were placed on the rate, amplitude, and energy content of 
gravitational waves.

\begin{figure}
\centering
\includegraphics[width=0.45\textwidth]{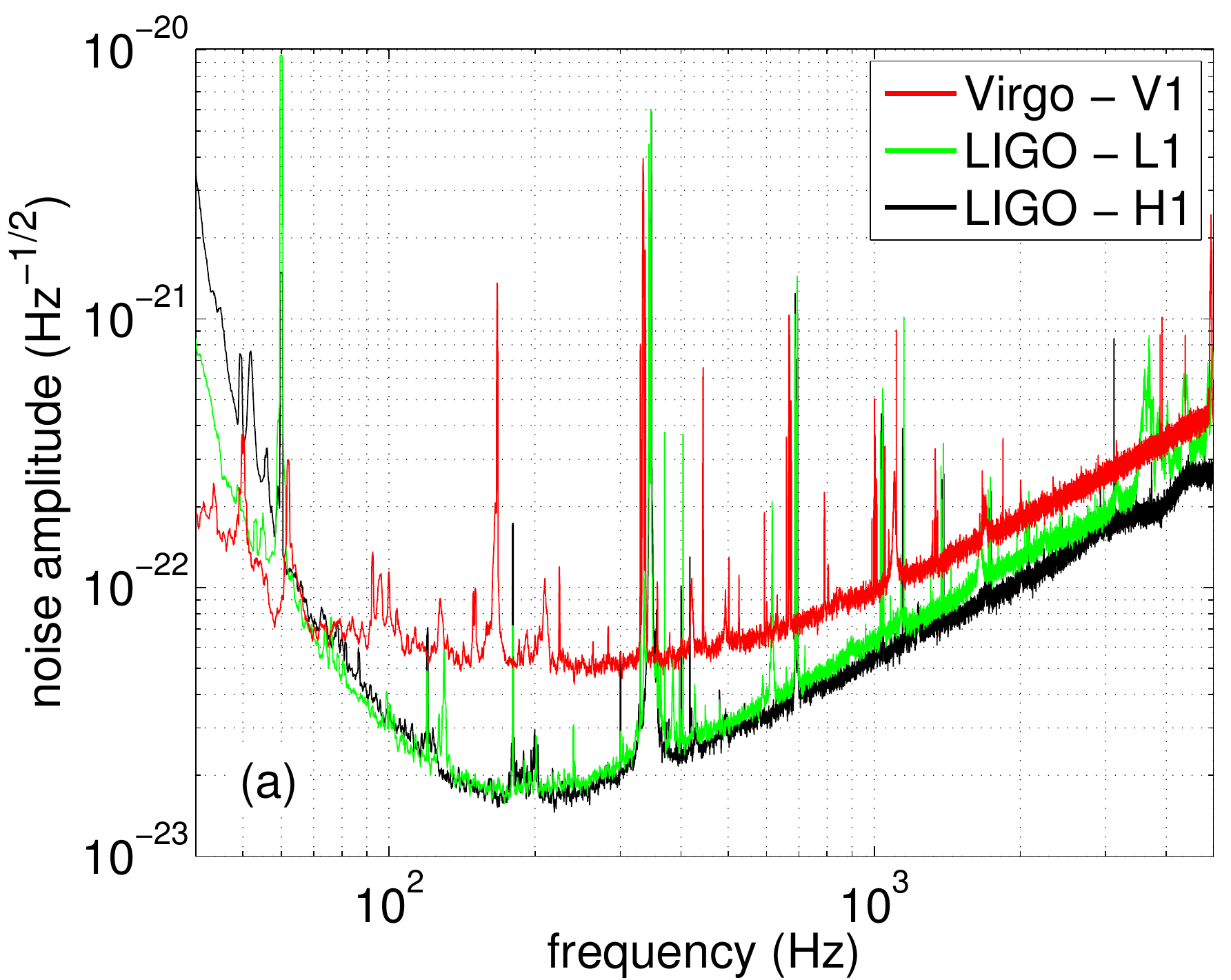}
\includegraphics[width=0.50\textwidth]{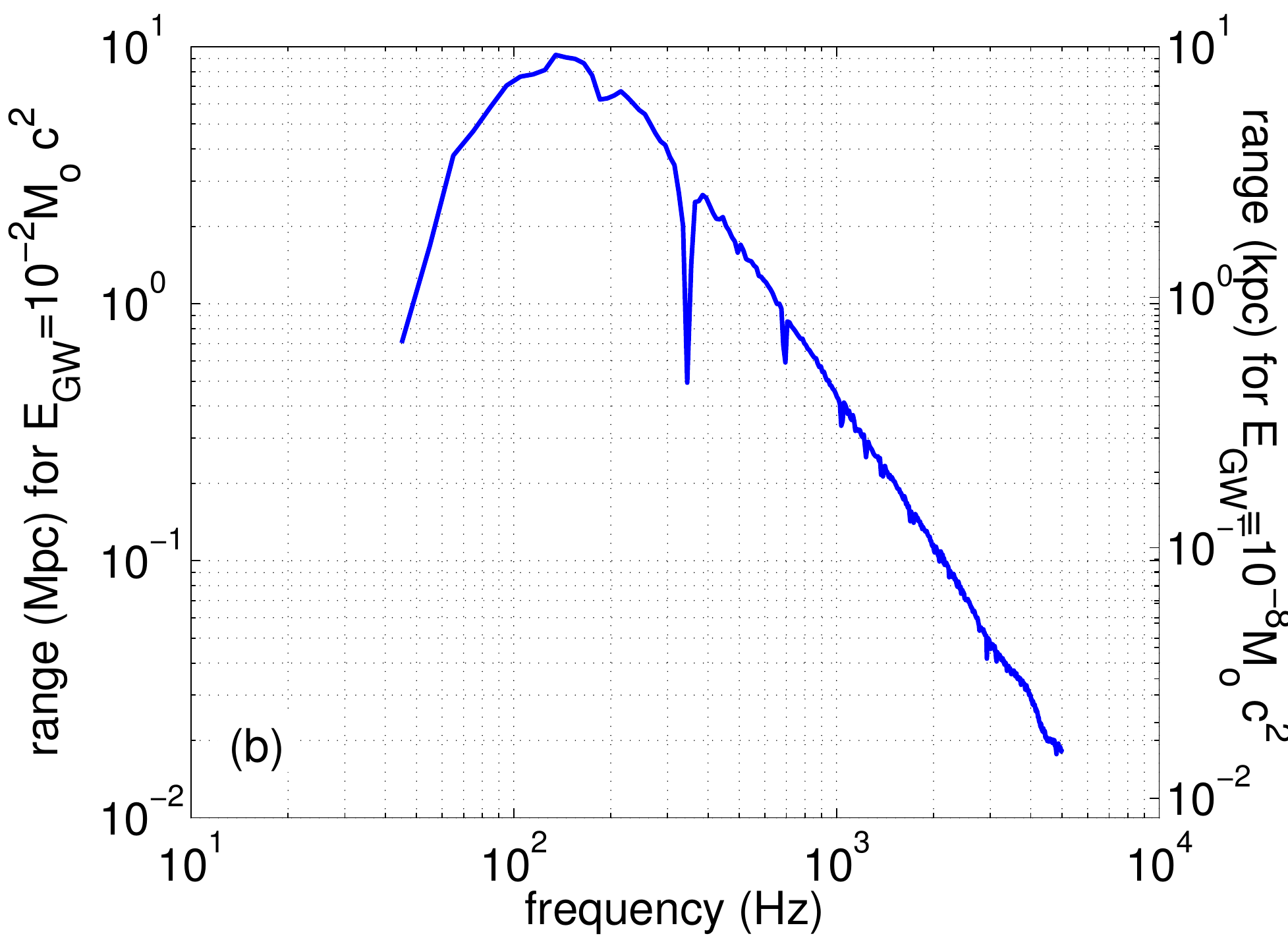}
\includegraphics[width=0.45\textwidth]{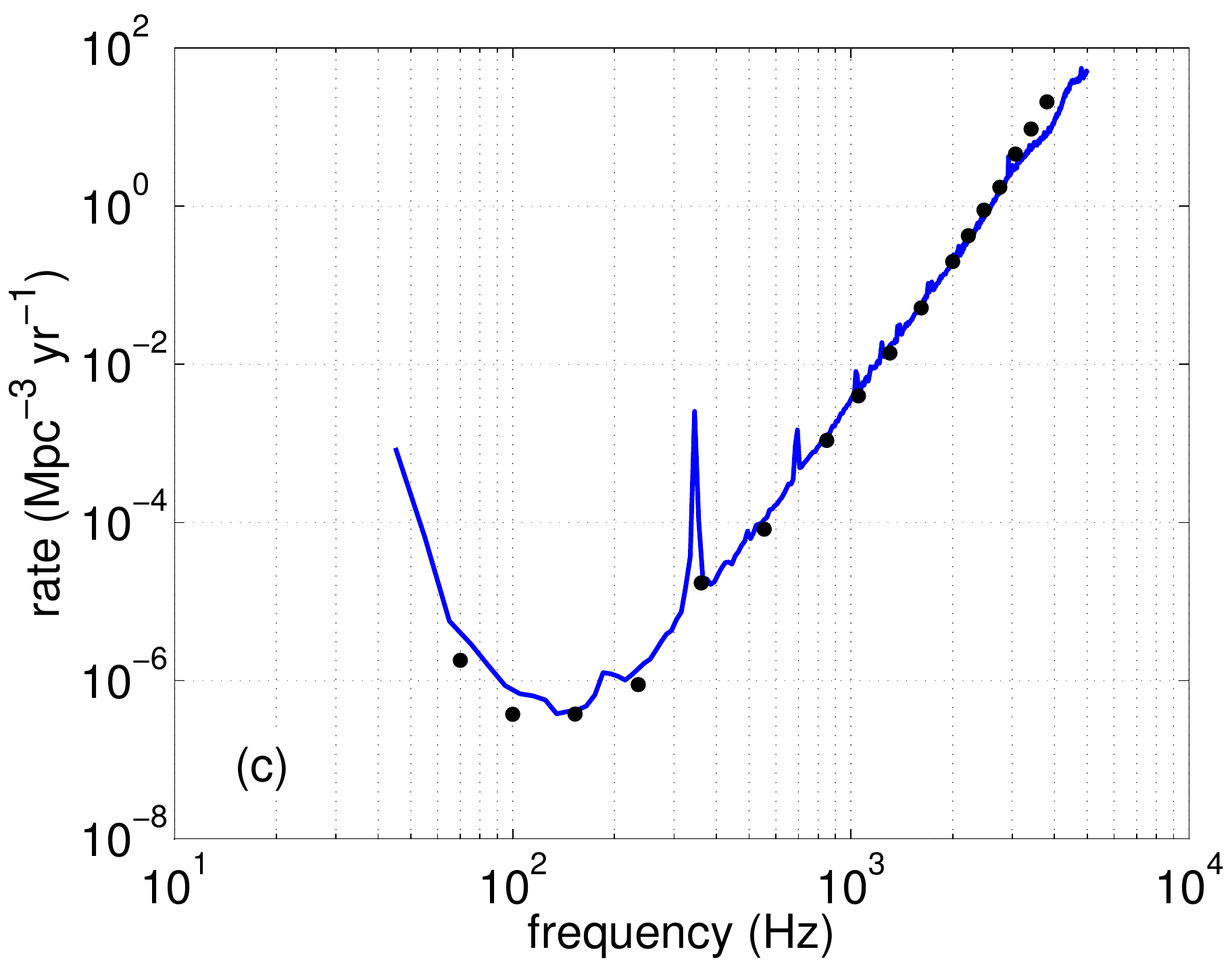}  
\includegraphics[width=0.45\textwidth]{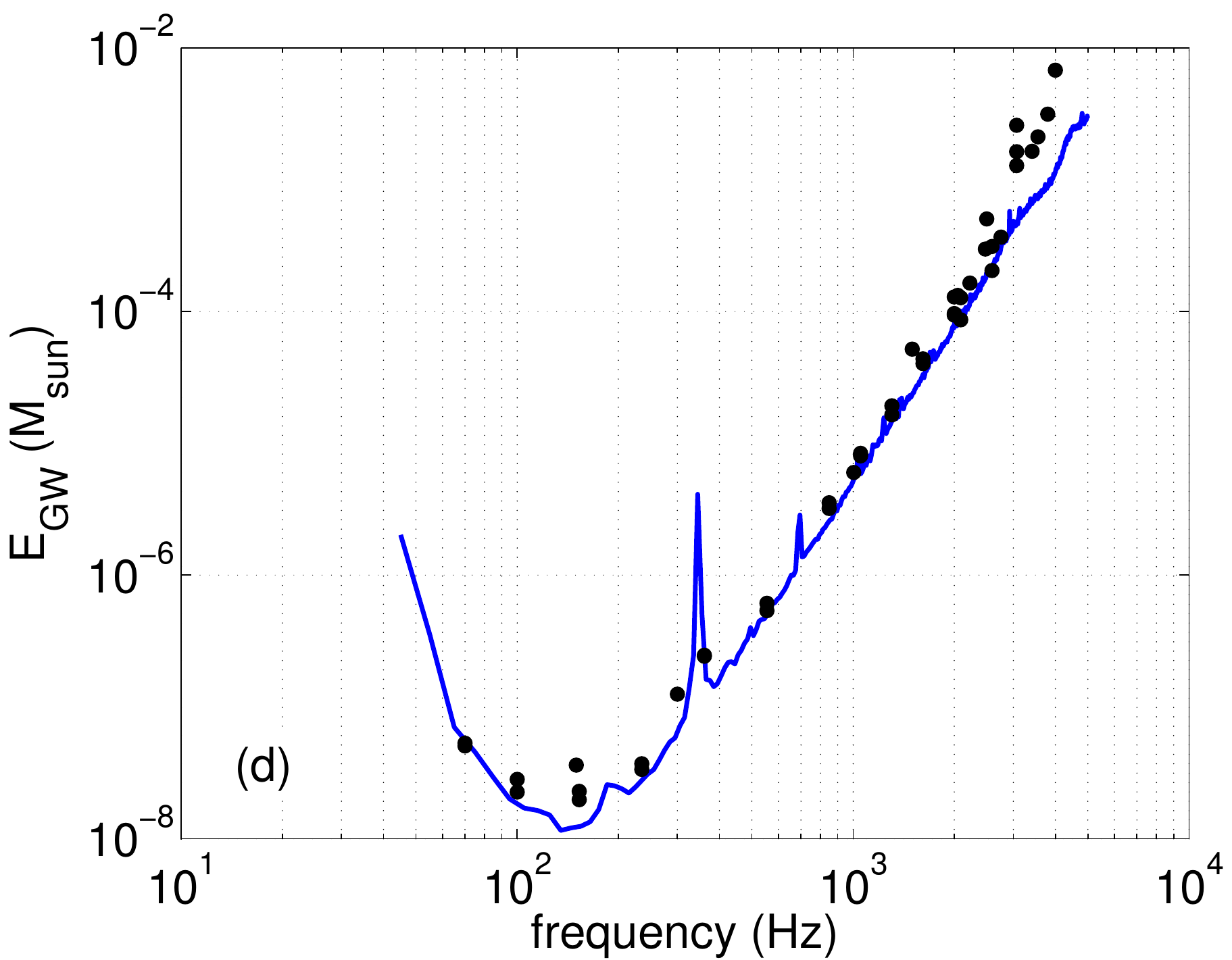}   
\caption{
(a) Example noise spectrum from each of the detectors in the 2009--10 
LIGO-Virgo science run.
(b) Effective range predicted by (\ref{eqn:range}) 
using $\rho_\mathrm{det}=20$ and the H1 noise curve smoothed 
to 10\,Hz resolution.  The left-hand (right-hand) scales assume 
$E_\mathrm{GW}=10^{-2}M_\odot c^2$ ($10^{-8}M_\odot c^2$).
(c) $E_\mathrm{GW}$ predicted by (\ref{eqn:range}) to be required 
for a source at a fixed distance of 10\,kpc to produce an expected SNR 
equal to $\rho_\mathrm{det}=20$.  The dots are the actual $E_\mathrm{GW}$ 
values for a variety of waveforms and the 2009--10 H1L1V1 network \cite{Ab_etal:12}.
(d) 90\% confidence rate density limits predicted by (\ref{eqn:rate}) 
for a homogeneous isotropic distribution of standard-candle 
sources with $E_\mathrm{GW}=1 M_\odot c^2$.
The dots are rate density limits for linearly polarized 
sine-Gaussian waveforms set by the combined 2005--07 and 2009--10 data 
sets \cite{Ab_etal:12}.
} 
\label{fig:plots}
\end{figure}

Figure~\ref{fig:plots}(a) shows an example noise spectrum from each of 
the detectors that participated in the 2009-10 run (data obtained from 
\cite{Ab_etal:12_publicdcc}).
The H1 detector had the lowest noise level across most of the search frequency 
band, so for convenience we use its noise spectrum $S(f)$ for our range 
calculations.  The other quantity required for defining the range is the 
SNR threshold $\rho_\mathrm{det}$, which is the threshold at which the detection 
efficiency is 50\%.  
Comparing the $h_\mathrm{rss}$ amplitude limits for 
linearly polarized sine-Gaussian bursts and unpolarized white-noise 
bursts (Tables II and IV of \cite{Ab_etal:12}, Table II of \cite{S5y2}, 
Fig.~3 of \cite{S5y1}) show that they correspond to 
$\rho_\mathrm{det}\simeq 20$ to 30 as measured against the H1 S6 noise 
spectrum\footnote{
The SNR threshold $\rho_\mathrm{det}$ can be estimated from 
the amplitude limit $h_\mathrm{rss}^{50\%}$ using (\ref{eqn:SNR2}, \ref{eqn:Theta}):  
$\rho_\mathrm{det} \simeq \Theta_\mathrm{rms} h_\mathrm{rss}^{50\%} / \sqrt{S(f_0)}$ where 
$\Theta_\mathrm{rms} =  (\langle F_+^2+F_\times^2 \rangle_{\theta,\phi,\psi})^{1/2} = \sqrt{2/5}$ 
for optimally oriented sources.
}, depending on the waveform, network, and data set. 
Since we expect the rate limits to be dominated by the most sensitive data 
(due to volume scaling), we select $\rho_\mathrm{det} = 20$ for our estimates.

Figure~\ref{fig:plots}(b) shows the effective range (\ref{eqn:range}) 
predicted assuming $\rho_\mathrm{det}=20$ and the H1 noise curve smoothed 
to 10\,Hz resolution.  The left-hand 
scale (Mpc) assumes $E_\mathrm{GW}=10^{-2}M_\odot c^2$, which is the 
approximate maximum gravitational-wave emission possible from long 
gamma-ray bursts under the most optimistic scenarios 
\cite{2002ApJ...579L..63D,Fryer,kobayashi-2003-585,Shibata03,Piro:2006ja,Corsi09,Romero10}.  
The right-hand scale (kpc) assumes $E_\mathrm{GW}=10^{-8}M_\odot c^2$, 
which is a typical energy emission in simulations of core-collapse 
supernovae \cite{Ott:2008wt,Kotake2011}.  The maximum ranges in the 
two cases are approximately 10\,Mpc (10\,kpc) for signal frequencies 
around 100\,Hz -- 200\,Hz, dropping to below 1\,Mpc (1\,kpc) by 
1000\,Hz.

Figure~\ref{fig:plots}(c) shows the $E_\mathrm{GW}$ predicted by 
(\ref{eqn:range}) to be required 
for a source at a fixed distance of 10\,kpc to produce an expected SNR 
equal to $\rho_\mathrm{det}=20$.  The dots are the actual $E_\mathrm{GW}$ 
values for a variety of waveforms and the 2009--10 H1L1V1 network, as 
reported in Fig.~7 of \cite{Ab_etal:12}.
Figure~\ref{fig:plots}(d) shows the 90\% confidence rate density limit 
(rate per unit volume) predicted by (\ref{eqn:rate}) 
for a homogeneous isotropic distribution of standard-candle 
sources with $E_\mathrm{GW}=1 M_\odot c^2$, assuming no detections in the 
2005--07 and 2009--10 searches (so that $\dot{N}\le2.3$ at 90\% confidence).
The dots are the approximate rate density limits for linearly polarized 
sine-Gaussian waveforms set by the combined 2005--07 and 2009--10 data 
sets (Fig.~6 of \cite{Ab_etal:12}).
Despite the fact that we use a single sample noise spectrum and SNR 
threshold to represent all networks and both science runs, the predicted 
limits are a reasonably good match to the measured limits in each case.


\section*{Acknowledgements}

The author would like to thank Eric Chassande-Mottin for motivating 
this investigation, and for his careful reading of and helpful 
suggestions on a previous draft.
This work was supported in part by STFC grants PP/F001096/1
and ST/J000345/1.
This draft has been assigned LIGO document number 
\href{https://dcc.ligo.org/LIGO-P1000041-v3}{{LIGO-P10}00041-v3}.

\newpage
\section*{References}

\bibliographystyle{unsrt}
\bibliography{References}

\begin{thebibliography}{10}

\bibitem{AnBrCrFl:01}
W.~G. Anderson, P.~R. Brady, J.~D.~E. Creighton, and \'E.~\'E. Flanagan.
\newblock Excess power statistic for detection of burst sources of
  gravitational radiation.
\newblock {\em Phys. Rev. D}, 63:042003, 2001.

\bibitem{FiCh:93}
Lee~Samuel Finn and David~F. Chernoff.
\newblock Observing binary inspiral in gravitational radiation: One
  interferometer.
\newblock {\em Phys. Rev. D}, 47(6):2198--2219, Mar 1993.

\bibitem{S5y1}
B.~P. {Abbott} et~al.
\newblock {Search for gravitational-wave bursts in the first year of the fifth
  LIGO science run}.
\newblock {\em Phys. Rev. D}, 80(10):102001, November 2009.

\bibitem{S5y2}
J.~{Abadie} et~al.
\newblock {All-sky search for gravitational-wave bursts in the first joint
  LIGO-GEO-Virgo run}.
\newblock {\em Phys. Rev. D}, 81(10):102001, May 2010.

\bibitem{Ab_etal:12}
J.~{Abadie} et~al.
\newblock {All-sky search for gravitational-wave bursts in the second joint
  LIGO-Virgo run}.
\newblock {\em Phys. Rev. D}, 85(12):122007, June 2012.

\bibitem{Su_etal:10}
P.~J. {Sutton}, G.~{Jones}, S.~{Chatterji}, P.~{Kalmus}, I.~{Leonor},
  S.~{Poprocki}, J.~{Rollins}, A.~{Searle}, L.~{Stein}, M.~{Tinto}, and
  M.~{Was}.
\newblock {X-Pipeline: an analysis package for autonomous gravitational-wave
  burst searches}.
\newblock {\em New Journal of Physics}, 12(5):053034, May 2010.

\bibitem{Ab_etal:12_publicdcc}
J.~{Abadie} et~al.
\newblock {All-sky search for gravitational-wave bursts in the second joint
  LIGO-Virgo run}.
\newblock Technical Report LIGO-P1100118-v24, LIGO Scientific Collaboration,
  Virgo Collaboration, 2012.
\newblock \url{https://dcc.ligo.org/LIGO-P1100118-v24/public}.

\bibitem{2002ApJ...579L..63D}
M.~B. {Davies}, A.~{King}, S.~{Rosswog}, and G.~{Wynn}.
\newblock {Gamma-Ray Bursts, Supernova Kicks, and Gravitational Radiation}.
\newblock {\em Astrophysical Journal}, 579:L63--L66, November 2002.

\bibitem{Fryer}
C.~L. Fryer, D.~E. Holz, and S.~A. Hughes.
\newblock {\em Astrophys. J.}, 565:430, 2002.

\bibitem{kobayashi-2003-585}
Shiho Kobayashi and Peter Meszaros.
\newblock Polarized gravitational waves from gamma-ray bursts.
\newblock {\em Astrophys. J.}, 585:L89, 2003.

\bibitem{Shibata03}
M.~Shibata, K.~Shigeyuki, and E.~Yoshiharu.
\newblock Dynamical bar-mode instability of differentially rotating stars:
  effects of equations of state and velocity profiles.
\newblock {\em Mon. Not. R. Astron. Soc.}, 343:619, 2003.

\bibitem{Piro:2006ja}
Anthony~L. {Piro} and Eric {Pfahl}.
\newblock {Fragmentation of Collapsar Disks and the Production of Gravitational
  Waves}.
\newblock {\em Astrophys. J.}, 658:1173--1176, April 2007.

\bibitem{Corsi09}
A.~Corsi and P.~Meszaros.
\newblock {GRB} afterglow plateaus and gravitational waves: Multi-messenger
  signature of a millisecond magnetar?
\newblock {\em Astrophys. J.}, 702:1171, 2009.

\bibitem{Romero10}
G.~E. Romero, M.~M. Reynoso, and H.~R. Christiansen.
\newblock Gravitational radiation from precessing accretion disks in gamma-ray
  bursts.
\newblock {\em Astron. Astrophys.}, 524:A4, 2010.

\bibitem{Ott:2008wt}
Christian~D. {Ott}.
\newblock {The gravitational-wave signature of core-collapse supernovae}.
\newblock {\em Class. Quantum Grav.}, 26(6):063001, March 2009.

\bibitem{Kotake2011}
K.~{Kotake}.
\newblock {Multiple physical elements to determine the gravitational-wave
  signatures of core-collapse supernovae}.
\newblock \url{http://arxiv.org/abs/1110.5107}.

\end{thebibliography}

\end{document}